\documentclass[floatfix,nofootinbib,superscriptaddress,twocolumn,pre,showpacs,showkeys,preprintnumbers]{revtex4-1}
\usepackage{amsmath,amssymb}
\usepackage{epsfig,graphics,color,calc,graphicx}
\usepackage{makeidx}
\usepackage{mathrsfs}

\newtheorem{theorem}{Theorem}
\newtheorem{condition}{Condition}

\newcommand{\qed}{\hfill $\Box$\bigskip}
\newcommand{\rr}[1]{{\normalfont\textrm{#1}}}

\newcommand{\bb}[1]{{\mathbb{#1}}}

\newcommand{\interk}{x^\rr{i}_\piccolo}
\newcommand{\interze}{x^\rr{i}_0}
\newcommand{\piccolo}{k}

\newlength{\pecettawidth}
\begin{document}
\title{Kink Localization under Asymmetric Double--Well Potential}

\author{Emilio N.M.\ Cirillo}
\email{emilio.cirillo@uniroma1.it}
\author{Nicoletta Ianiro}
\email{nicoletta.ianiro@uniroma1.it}
\affiliation{Dipartimento di Scienze di Base e Applicate per l'Ingegneria, 
             Sapienza Universit\`a di Roma, 
             via A.\ Scarpa 16, I--00161, Roma, Italy.}

\author{Giulio Sciarra}
\email{giulio.sciarra@uniroma1.it}
\affiliation{Dipartimento di Ingegneria Chimica Materiali Ambiente,
             Sapienza Universit\`a di Roma, 
             via Eudossiana 18, I--00184 Roma, Italy}


\begin{abstract}
We study diffuse phase interfaces under asymmetric double--well 
potential energies with degenerate minima and demonstrate that 
the limiting sharp profile, for small interface energy cost, 
on a 
finite space interval is in general not symmetric and its position 
depends exclusively on the second derivatives of the 
potential energy at the two minima (phases).
We discuss an application of the general result 
to porous media 
in the regime of solid--fluid segregation under an 
applied pressure and describe the interface between 
a fluid--rich and 
a fluid--poor phase. 
\end{abstract}





\maketitle

\section{Introduction}
\label{s:introduzione}
Quench a system from the homogeneous phase into a
broken--symmetry one (think, e.g., to a ferromagnet or to 
a gas abruptly
cooled below their critical temperature). 
The two phases have to separate and the 
process can be described via a local field $u$ 
(local magnetization in ferromagnets, density in liquid--vapor
systems, concentration in alloys, etc.)
on the physical space $\Omega\subset\bb{R}^d$.

Very well known 
models \cite{Bray} 
for the description of the field evolution are 
the Allen--Cahn
and the Cahn--Hilliard equations.
With suitable boundary conditions 
the former is an appropriate equation when the 
order parameter, i.e., the integral over $\Omega$ of the 
field $u$, is not conserved, while the latter applies in the 
conserved case.
On mathematical grounds, 
these equations can be thought as the gradient equation 
$\partial u/\partial t=-\delta F/\delta u$, 
in a suitable Hilbert space \cite{AF,Fife,CIS2011b}, 
for the \textit{Landau energy functional}
\begin{equation}
\label{funzionale}
F[u]
=
\int_\Omega\bigg[\frac{1}{2}\piccolo^2\|\nabla u\|^2+V(u)\bigg]\,\rr{d}x
\end{equation}
with $\piccolo>0$ and the \textit{potential energy}
$V$ a double well positive regular function with degenerate zero value 
absolute minima in $a,b$, called \textit{phases} of the system.
More precisely, 
if no constraint to the order parameter is imposed, 
it is possible to compute the gradient of the Landau functional 
in the Hilbert space $L^2(\Omega)$
to get the Allen--Cahn equation \cite[equations (8) and (12)]{Fife}. 
When the integral of the field $u$ 
is assumed to be constant throughout the evolution, computing 
the gradient in the $L^2(\Omega)$ results into a non--local 
evolution equation, while by using the Hilbert space 
$H^{-1}(\Omega)$ the Cahn--Hilliard equation is found
\cite[equation (91) and the comment below equation (90)]{Fife}.

In this paper we discuss some properties of the stationary profile 
connecting the two phases $a$ and $b$, namely, the 
profile which is reached by the system at infinite time. 
More precisely, 
we explore the properties of the stationary 
profile connecting the two phases when the 
double well potential energy is asymmetric. 
The equations for the interface profile are 
the Euler--Lagrange equations
for the stationary points of the energy functional (\ref{funzionale}).
These kind of problems are often referred to in the literature 
as \emph{gradient} or \emph{diffuse interface} problems \cite{Modica}.

We shall consider a one--field $u$ system 
described by a potential energy $V$ 
as above. 
More precisely we shall assume, throughout the paper, that 
the following condition is satisfied. 

\begin{condition}
\label{t:cond}
The function $V:\bb{R}\to\bb{R}$ is a positive $C^2(\bb{R})$ function with 
two single isolated local minima $a$ and $b$ (assume $a<b$) and such that 
$V(a)=V(b)=0$, 
$V''(a)>0$, and $V''(b)>0$.
\end{condition}

Under this assumption, we shall approach the phase interface problem
in dimension one, on a finite interval with Dirichlet boundary
conditions fixing the values of the two phases at the extreme points
of the interval to $a$ and $b$, respectively. We prove that, for small 
$\piccolo$,
the position of the interface depends exclusively on the second
derivatives of the potential energy $V$ at the minima $a$ and $b$. In 
particular, the
statement is valid for the stationary solutions of the 
Allen-–Cahn equation at imposed zero
chemical potential and for those of the Cahn-–Hilliard equation at imposed
zero Lagrange multiplier (for mass conservation).

Note that 
even if $V$ is not symmetric,
the interface falls in the middle point 
of the space interval
provided 
the two second derivatives are mutually equal.



The interest to study the interface problem in the case of an 
asymmetric double--well potential energy is due to the fact that 
they appear naturally in different 
physical situations. For example, 
in \cite{CIS2009,CIS2010,CIS2011,CIS2011b}
we introduced 
a model describing a pressure guided 
transition between a fluid--rich and a fluid--poor 
phase
and based on 
an asymmetric double--well potential energy. 
In this theory two order parameters $m$ and $\varepsilon$ are introduced, 
having respectively 
the physical meaning of fluid density and solid strain. 
In \cite{CIS2010,CIS2011b} the interface separating the two coexisting 
phase has been studied and it has been shown that its localization
properties, due to the asymmetry of the potential energy, are not 
trivial. 


Apart from the application mentioned above, due to its general 
character, our result is of interest for any situation in which 
diffuse interfaces are relevant. In particular when approximate 
computations or numerical simulations 
are performed in the regime of small interface energy cost, our 
rigorous prediction could result valuable in testing 
the soundness of the results. We remark that in this kind 
of problems numerical computations become particularly difficult 
when the interface energy cost is small (say $k\sim10^{-2}$ with 
our parametrization). 

The paper is organized as follows. In Section~\ref{s:risultati} we 
state our general results. In Section~\ref{s:porosi} we discuss in detail 
their application to the porous media segregation problem. 
In Section~\ref{s:motore} we shall verify our result in 
some cases physically relevant by performing numerical computations.
In Section~\ref{s:dimostrazione}, finally, we prove the results stated in 
Section~\ref{s:risultati}.

\section{Results}
\label{s:risultati}
We approach the problem in dimension one. 
Recalling $\piccolo$ is a positive constant,
the Euler--Lagrange equations for the energy functional 
(\ref{funzionale}) 
reads as the Dirichlet boundary value problem 
\begin{equation}
\label{dir00}
\piccolo^2 u_{xx}=V'(u)
\;\textrm{ with }\;
u(0)=a
\;\textrm{ and }\;
u(\ell)=b
\end{equation}
for the field $u(x)$, 
$x\in[0,\ell]$, for some $\ell>0$.

From the physical point of view the problem (\ref{dir00}) is that 
of finding the stationary profile connecting the two phases $a$ and $b$ on the 
finite space interval $[0,\ell]$.
We remark that the above problem (\ref{dir00}) is found when 
looking for the stationary solutions of 
the Allen-–Cahn equation at imposed zero
chemical potential or of the Cahn-–Hilliard equation at imposed zero
Lagrange multiplier (for mass conservation) with proper boundary
conditions.

Before stating our results, we note that 
by exploiting the one--dimensionality of the model
a phase space analysis \cite{CI} proves that 
the problem (\ref{dir00}) has a unique solution implicitly 
given by the integral 
\begin{equation}
\label{dir01}
\int_a^u\frac{\piccolo\,\rr{d}s}
             {\sqrt{2[E_\piccolo+V(s)]}}
=
x
\end{equation}
where for any $\piccolo>0$ we have defined implicitly 
the energy level $E_\piccolo$
by the equation 
\begin{equation}
\label{disc02}
\int_a^b\frac{\piccolo\,\rr{d}s}
             {\sqrt{2[E_\piccolo+V(s)]}}
=
\ell
\end{equation}
namely, 
the integral in (\ref{dir01}) with $u=b$ and $x=\ell$.

\begin{figure}
\begin{picture}(100,100)
\put(-140,-360)
{
\resizebox{24cm}{!}{\rotatebox{0}{\includegraphics{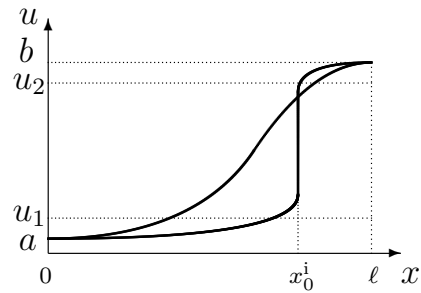}}}
}
\end{picture}
\vskip 1. cm
\caption{Schematic description of the results in Theorems~\ref{t:disc} 
and \ref{t:interfaccia00}.
The thick lines are the profiles at two finite values of $\piccolo$; 
the steepest one corresponds to the smallest value of $\piccolo$.
}
\label{f:schematico}
\end{figure}

We note that 
the solution of (\ref{dir01}) is a kink
connecting on the interval $[0,\ell]$ the phase $a$
to the phase $b$. At small $\piccolo$ the energy cost
associated with the gradient of $u$, see (\ref{funzionale}),
is small, so that the interface width is of order $\piccolo$
and it is localized somewhere
in the interval $[0,\ell]$.
This remark is made rigorous 
in the next classical theorem, where we state that in the limit 
$\piccolo\to0$ the interface 
tends to a discontinuous kink connecting the two phases (see, 
also, figure~\ref{f:schematico}).

\begin{figure*}
\begin{picture}(100,100)
\put(-220,-360)
{
\resizebox{24cm}{!}{\rotatebox{0}{\includegraphics{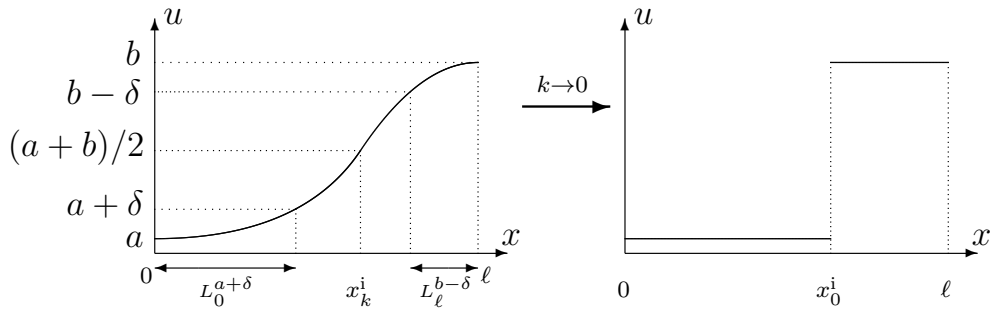}}}
}
\end{picture}  
\vskip 1. cm
\caption{Schematic description of the heuristic interpretation of 
equation (\ref{interfaccia00}).
The thick line is the profile at finite $k$, whereas the thin 
discontinuous line is its $k\to0$ limit.
}
\label{f:congettura}
\end{figure*}

\begin{theorem}
\label{t:disc}
Assume Condition~\ref{t:cond} is satisfied.
Then, for any 
$u_1,u_2\in(a,b)$ such that $u_1<u_2$ 
\begin{equation}
\label{disc}
\lim_{\piccolo\to0}
\int_{u_1}^{u_2}
\frac{\piccolo\,\rr{d}s}
             {\sqrt{2[E_\piccolo+V(s)]}}
=0
\end{equation}
\end{theorem}

The meaning of the theorem above is illustrated in figure~\ref{f:schematico}.
For any choice of 
$a<u_1<u_2<b$, by choosing $\piccolo$ small enough the 
distance between the two points where the kink attains the values 
$u_1$ and $u_2$ respectively can be made smaller than any fixed positive 
number. 

The theorem above suggests that for any 
$\piccolo>0$, it is meaningful to define the \textit{interface
position} $\interk$
as that (unique) point where the profile attains
the value $(a+b)/2$, that is to say
\begin{equation}
\label{interk}
\interk
=
\int_a^{\frac{a+b}{2}}
\frac{\piccolo\,\rr{d}s}
             {\sqrt{2[E_\piccolo+V(s)]}}
\end{equation}
Since for $\piccolo\to0$ 
the profile 
tends to a step function, 
different definitions, equivalent in this limit,
are possible for the interface position;
e.g., 
the point where the profile 
equals the value of field corresponding to the local maximum
of the potential energy.

The following theorem gives the limiting behavior of the 
interface position for $\piccolo\to0$. 

\begin{theorem}
\label{t:interfaccia00}
Assume Condition~\ref{t:cond} is satisfied.
Then 
\begin{equation}
\label{interfaccia00}
\lim_{\piccolo\to0}
 [\sqrt{V''(a)}\,\interk-\sqrt{V''(b)}\,(\ell-\interk)]=0
\end{equation}
\end{theorem}

In words, the theorem states that 
in the limit $\piccolo\to0$ the distance of the 
interface from the boundary points is proportional to the inverse 
of the square root of the second derivative of the potential 
energy evaluated at the corresponding minimum of $V$.
Note that from (\ref{interfaccia00}) it follows immediately that 
the interface position $\interk$ 
for $\piccolo\to0$ tends to 
\begin{equation}
\label{interfaccia01}
\interze=\frac{\ell\sqrt{V''(b)}}
              {\sqrt{V''(a)}+\sqrt{V''(b)}}
\end{equation}
The limiting interface position $x^\rr{i}_0$ 
depends only on the second derivatives of the potential energy evaluated 
at the 
minima. In particular if $V''(a)=V''(b)$ then $x^\rr{i}_0=\ell/2$, 
that is to say, the interface is located at the middle point of the 
interval $[0,\ell]$. 

The proof (see Section~\ref{s:dimostrazione})
of (\ref{interfaccia00}) is based on a direct evaluation 
of the integral (\ref{interk}).
Since $E_\piccolo$ is small for $\piccolo\to0$ the integrand diverges 
in $a$. The integral can be thus estimated by expanding 
in Taylor formula to the second order the potential energy $V$. 

Now, 
we give two heuristic arguments 
supporting the statement of the theorem.
In Section~\ref{s:motore}, on the other hand, 
we shall verify the result with some 
numerical computations in two cases of interest.

Assuming $\piccolo$ small, the kink is close 
to a step function. By phase space techniques it follows
$E_\piccolo\approx0$. Fix $\eta>0$ small 
and $\eta<\delta\ll b-a$. By (\ref{dir01})
the distance between the points where the profile equals
$a+\eta$ and $a+\delta$ is
\begin{displaymath}  
\int_{a+\eta}^{a+\delta}\frac{\piccolo\,\rr{d}s}
                             {\sqrt{2[E_\piccolo+V(s)]}}
\approx
\int_{a+\eta}^{a+\delta}\frac{\piccolo\,\rr{d}s}
                             {\sqrt{V''(a)}(u-a)}
\end{displaymath}  
where we have neglected $E_\piccolo$ and, noted 
that the integral is extend to a very small interval close to $a$, 
we expanded $V$ in second order Taylor formula.
Since $\eta$ is small, we have that the distance $L_0^{a+\delta}$ 
from $0$ of 
the point where the value $a+\delta$ is attained by the profile is 
approximatively given by 
$L_0^{a+\delta}
\approx
(\piccolo/\sqrt{V''(a)})\log(\delta/\eta)$.

By performing a similar computation in the neighborhood
of the boundary point $\ell$
and with obvious notation we get 
$
L_\ell^{b-\delta}
\approx
(\piccolo/\sqrt{V''(b)})\log(\delta/\eta)$.
The last two formulas allow to conjecture the validity of 
(\ref{interfaccia00}), see also figure~\ref{f:congettura}.

The second heuristic argument is based on the interpretation of 
(\ref{dir00}) as 
the equation describing the motion of a particle of mass $\piccolo^2$ 
under the potential energy $-V$. In this language $u$ is the position 
of the particle and $x$ is the time. In the remaining part of this section, 
thus, we shall address to $u(x)$ as to the position of the particle 
at time $x$ and, for this reason, derivatives with respect to $x$ 
will be denoted by dots. 

In this context, solving (\ref{dir00}) means looking for 
the motion started at $a$ with positive initial velocity $v_0$ such that 
at time $\ell$ the position $b$ is reached. 
The conservation of the total mechanical energy $\piccolo^2\dot{u}^2/2-V(u)$
implies that the motion will reach the point $b$ with velocity $v_0$. 
The interface position $\interk$ is then the time at which the 
particle reaches the position $(a+b)/2$.

We can 
describe such a motion by linearizing the equation of motion
in a neighborhood of the two unstable equilibrium points 
$a$ and $b$. We get 
\begin{displaymath}
\piccolo^2\ddot{u}=V''(a)(u-a)
\;\textrm{ and }\;
\piccolo^2\ddot{u}=V''(b)(u-b)
\end{displaymath}
respectively.
We then solve the two equations with the initial conditions
$(u(0),\dot{u}(0))=(a,v_0)$
and
$(u(0),\dot{u}(0))=(b,-v_0)$,
respectively, and get 
\begin{displaymath}
u_a(x)=a+\frac{\piccolo v_0}{\sqrt{V''(a)}}
               \sinh\Big(\frac{\sqrt{V''(a)}}{\piccolo}x\Big)
\end{displaymath}
for the motion started at $a$ and 
\begin{displaymath}
u_b(x)=b-\frac{\piccolo v_0}{\sqrt{V''(b)}}
               \sinh\Big(\frac{\sqrt{V''(b)}}{\piccolo}x\Big)
\end{displaymath}
for the motion that started at $b$.

Recalling that $\interk$ is the time at which the particle 
is at $(a+b)/2$, by the two equation above we get 
\begin{displaymath}
\interk=\frac{\piccolo}{\sqrt{V''(a)}}\,
       \mathrm{arcsinh}
          \Big(\frac{b-a}{2}\frac{\sqrt{V''(a)}}{\piccolo v_0}\Big)
\end{displaymath}
and 
\begin{displaymath}
\ell-\interk=\frac{\piccolo}{\sqrt{V''(b)}}\,
       \mathrm{arcsinh}
          \Big(\frac{b-a}{2}\frac{\sqrt{V''(b)}}{\piccolo v_0}\Big)
\end{displaymath}

Recalling that $\mathrm{arcsinh}(x)=\log(x+\sqrt{x^2+1})$, 
with simple algebra we get 
\begin{widetext}
\begin{equation}
\label{heur00}
\sqrt{V''(a)}\interk-\sqrt{V''(b)}(\ell-\interk)
=
\piccolo
\bigg[
\log\frac
       {\sqrt{V''(a)}(b-a)+\sqrt{V''(a)(b-a)^2+4\piccolo^2v_0^2}}
       {\sqrt{V''(b)}(b-a)+\sqrt{V''(b)(b-a)^2+4\piccolo^2v_0^2}}
\bigg]
\end{equation}
\end{widetext}
which suggests the validity of (\ref{interfaccia01}) since the 
right hand side tends to zero for $\piccolo\to0$. Note that 
$v_0$ does not diverge 
for $\piccolo\to0$.

\section{Phase interface in porous media}
\label{s:porosi}
We discuss, now, in detail the application of the above results 
to the study of transitions 
from a fluid--poor towards a fluid--rich phase 
in porous media under consolidation.
The model we study here is supposed to explain
the behavior of the system in consolidation regime, namely, when
an external pressure is applied on the solid component. The idea
is that of explaining the existence of two phases, differing in liquid
content, due to the modification in the solid structure caused by
the external pressure. For this reason, as in the original Biot
theory, in this model no chemical potential contribution is considered.

\begin{figure}
\begin{picture}(100,110)
\put(-70,-30)
{
\resizebox{14cm}{!}{\rotatebox{0}{\includegraphics{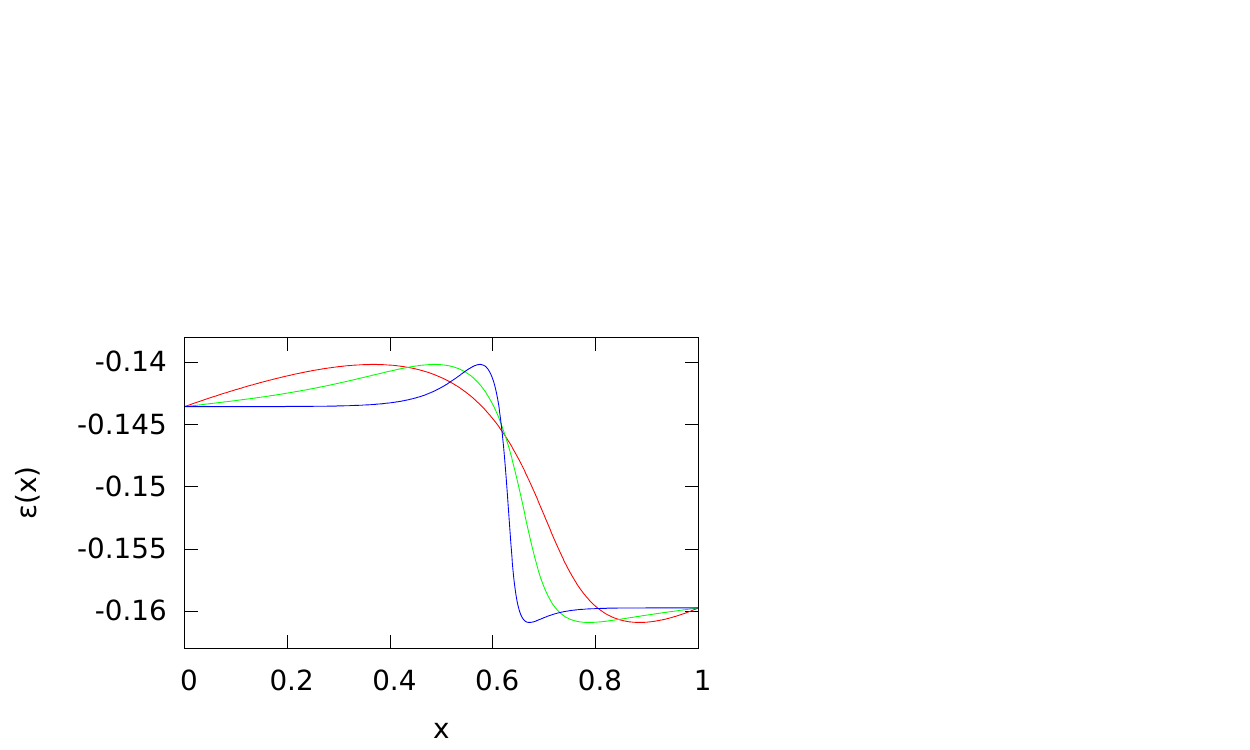}}}
}
\end{picture}  
\vskip 1. cm
\caption{Kink profiles for the strain field $\varepsilon$ 
for the potential energy 
(\ref{kappa})--(\ref{secondo020}) with
$k_1=k$, $k_2=k$, $k_3=k$, $k=10^{-1},10^{-2},10^{-3}$, 
$\alpha=100$, $a=1/2$, and $b=1$.}
\label{f:orbe1}
\end{figure}

\begin{figure}
\begin{picture}(100,110)
\put(-70,-30)
{
\resizebox{14cm}{!}{\rotatebox{0}{\includegraphics{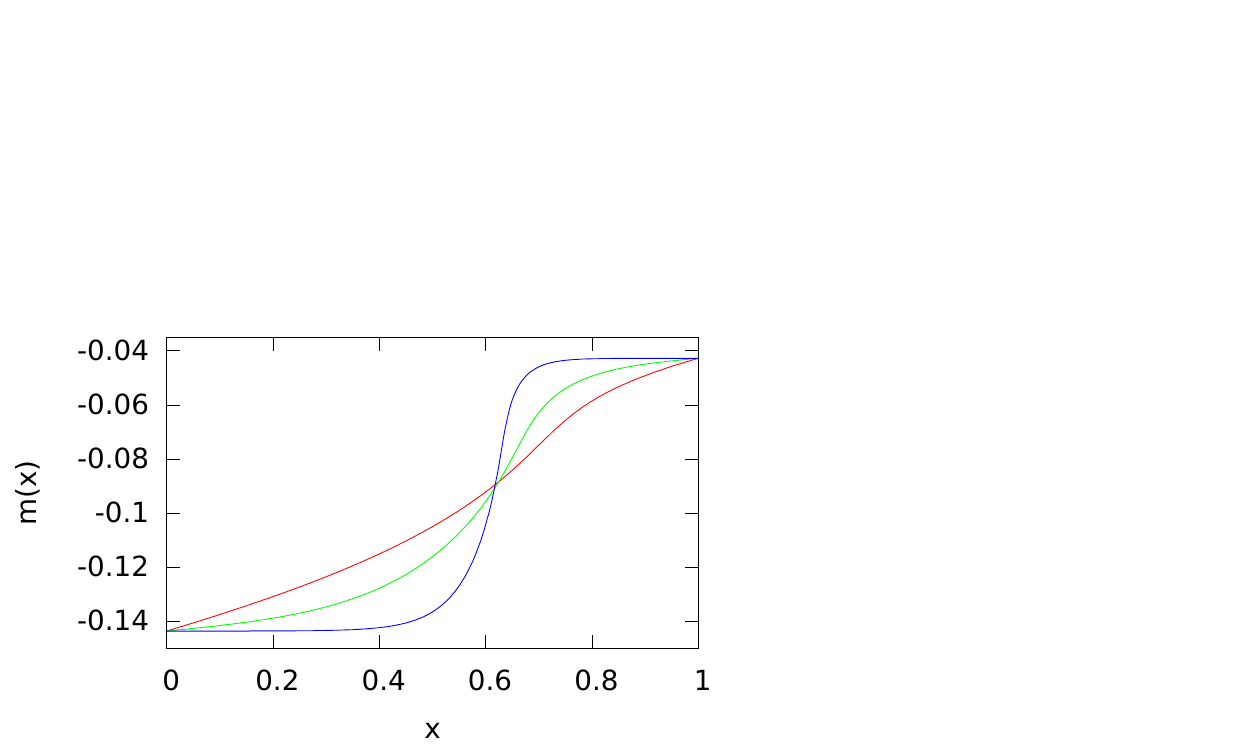}}}
}
\end{picture}  
\vskip 1. cm
\caption{Kink profiles for the fluid density field $m$ for the 
same parameters as in figure~\ref{f:orbe1}.}
\label{f:orbm1}
\end{figure}

\begin{figure}
\begin{picture}(100,110)
\put(-70,-30)
{
\resizebox{14cm}{!}{\rotatebox{0}{\includegraphics{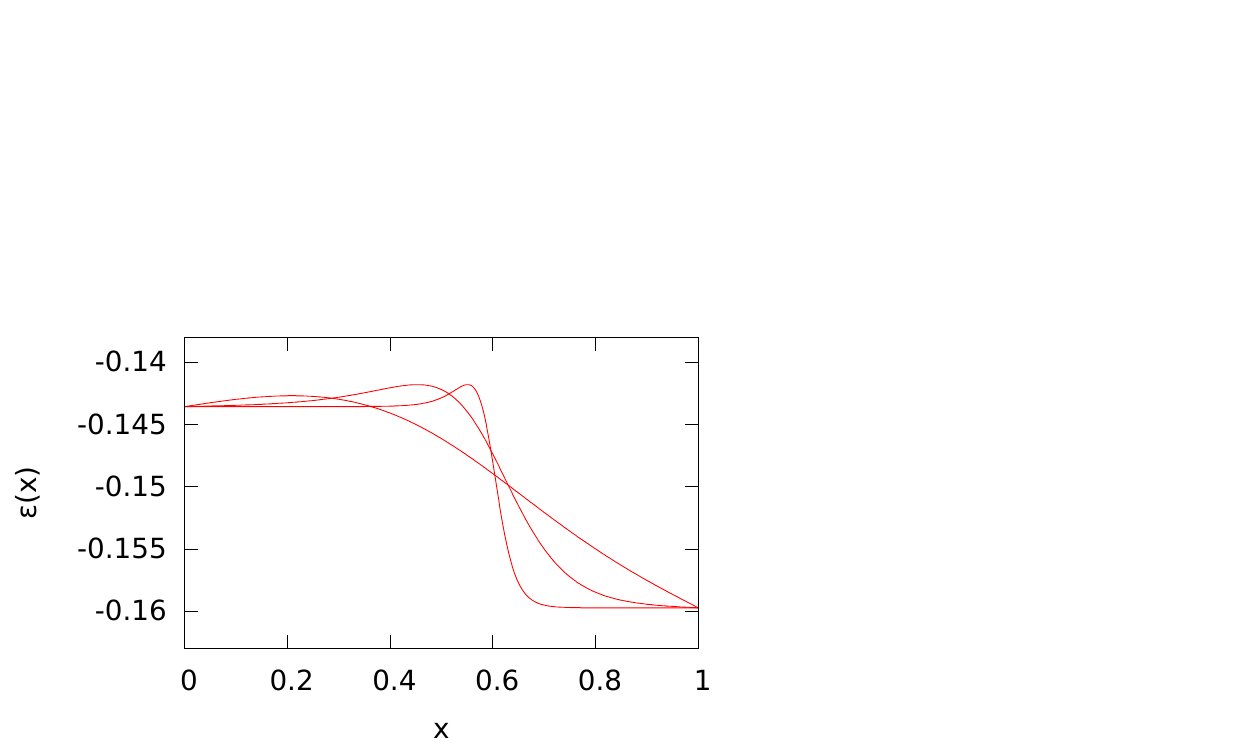}}}
}
\end{picture}  
\vskip 1. cm
\caption{Kink profiles for the strain field $\varepsilon$ 
for the potential energy 
(\ref{kappa})--(\ref{secondo020}) with
$k_1=k$, $k_2=k/2$, $k_3=k/3$, $k=10^{-1},10^{-2},10^{-3}$, 
$\alpha=100$, $a=1/2$, and $b=1$.}
\label{f:orbe2}
\end{figure}

\begin{figure}
\begin{picture}(100,110)
\put(-70,-30)
{
\resizebox{14cm}{!}{\rotatebox{0}{\includegraphics{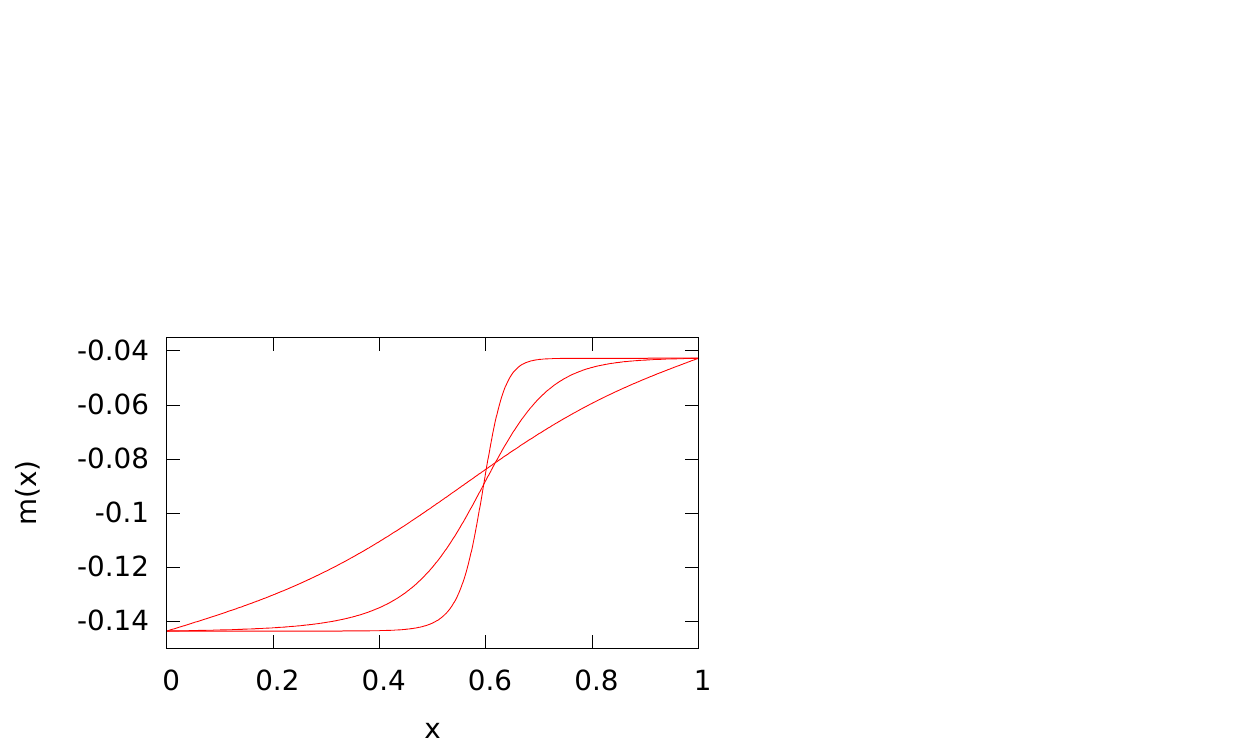}}}
}
\end{picture}  
\vskip 1. cm
\caption{Kink profiles for the fluid density field $m$ for the 
same parameters as in figure~\ref{f:orbe2}.}
\label{f:orbm2}
\end{figure}

In \cite{CIS2010,CIS2011b} the interface separating the two coexisting 
phases has been studied numerically and it has been shown that its localization
properties, due to the asymmetry of the potential energy, are not 
trivial. In figures~\ref{f:orbe1}--\ref{f:orbm2} profiles 
for the solid strain $\varepsilon$ and the 
fluid density $m$ fields are shown. It is immediate 
to note that those kinks tend, for the interface energy cost tending to zero, 
to a not symmetric sharp interface between the two phases.  

More precisely, in this model three interface cost parameters, $k_1$,
$k_2$, and $k_3$, are introduced (see equation (\ref{kappa}) below).
These constants weights $(\varepsilon')^2$, $\varepsilon'm'$, and 
$(m')^2$, respectively. 
The numerical simulations suggest that when $k_1,k_2,k_3\to0$ with 
mutual ratios kept constant, the interface tends to a definite position 
not depending on these ratios. 
The $\varepsilon$ and the $m$ profiles are shown 
in the cases 
$k_1=k_2=k_3$
in figures~\ref{f:orbe1} and \ref{f:orbm1}
and
$k_1=2k_2=3k_3$
in figures~\ref{f:orbe2} and \ref{f:orbm2},
respectively. 

In Section~\ref{s:risultati} we have proven results concerning the localization 
properties of a kink connecting two coexisting phases 
in a quite general one--field 
one--dimensional setup.
As a straightforward application, we shall get the position of the 
limiting interface for the poromechanics model described above with
the second gradient coefficients such that 
$k_1k_3-k_2^2=0$; in this particular case, indeed, the two--field 
model will reduce to a one--field one. 

For the complete description of the poromechanics setup 
we refer the readers to \cite{CIS2010,CIS2011,CIS2011b,SIIM}.
We recall that two fields $m,\varepsilon:[0,\ell]\to\bb{R}$ are 
introduced having the physical meaning of deviation from a reference 
value of the fluid density and solid strain, respectively. 
The stationary state of the system is described by the 
Euler--Lagrange equations, with suitable boundary conditions, 
for the variational problem associated with the Landau 
energy functional 
\begin{equation}
\label{funzionale02}
L[m,\varepsilon]
=
\int_0^\ell
 [K(m',\varepsilon')+\Psi(m,\varepsilon)]
 \,\rr{d}x
\end{equation}
where $K$ and $\Psi$ are, respectively, called the 
\textit{second} and \textit{first} gradient potential energy. 

The poromechanics model we study in this paper \cite{CIS2011} 
is defined by choosing the second and the first gradient potential 
energies as follows. We let 
\begin{equation}
\label{kappa}
K(m',\varepsilon')
=
\frac{1}{2}
 [
  k_1(\varepsilon')^2+2k_2\varepsilon'm'+k_3(m')^2]
\end{equation}
where $k_1,k_3>0$ and $k_2\in\bb{R}$ are such that $k_1k_3-k_2^2\ge0$ and 
the prime denotes the space derivative. 
Moreover we set 
\begin{equation}
\label{psi01}
\Psi(m,\varepsilon,p)
 \!=\!
 \frac{\alpha}{12}m^2(3m^2\!-8b\varepsilon m+6b^2\varepsilon^2)
 +\!
 \Psi_\rr{B}(m,\varepsilon,p)
\end{equation}
where 
\begin{equation}
\label{secondo020}
\Psi_\rr{B}(m,\varepsilon;p)=
 p\varepsilon+\frac{1}{2}\varepsilon^2+\frac{1}{2}a(m-b\varepsilon)^2
\end{equation}
is the Biot potential energy density~\cite{biot01},
$a>0$ is the ratio between the fluid and the solid rigidity, 
$b>0$ is a coupling between the fluid and the solid component, 
$p>0$ is the external pressure,
and
$\alpha>0$ is a material parameter responsible for the showing 
up of the additional equilibrium.
We remark that the condition 
$k_1k_3-k_2^2\ge0$ ensures that the second gradient 
part $K$ of the overall potential energy density is convex.
Under this assumption there exists a minimizer 
for the action functional (\ref{funzionale02})
on a bounded domain \cite{Evans}.

In~\cite{CIS2011,CIS2009}
we have studied the minima of the potential energy (\ref{psi01})
and shown that they can be interpreted as the homogeneous 
\textit{phases} of the system. We have 
proven that there exists a pressure $p_\rr{c}$, called 
\textit{critical pressure}, such that for any $p\in[0,p_\rr{c})$ there 
exists a single phase 
$(m_\rr{s}(p),\varepsilon_\rr{s}(p))$, 
called 
the \textit{standard phase}, which is very similar to the usual 
solution of the Biot model. For $p> p_\rr{c}$ a second phase 
$(m_\rr{f}(p),\varepsilon_\rr{f}(p))$, richer in fluid with respect to the 
standard phase and hence called \textit{fluid--rich} phase, appears.
It has also been shown that for any $p>0$ the point 
$(m_\rr{s}(p),\varepsilon_\rr{s}(p))$ 
is a local minimum of the two variable potential 
energy $\Psi(m,\varepsilon)$ with $p$ fixed, while
$(m_\rr{f}(p),\varepsilon_\rr{f}(p))$ 
is a local minimum for $p>p_\rr{c}$ and 
a saddle point for $p=p_\rr{c}$.

Moreover, 
in \cite{CIS2010} 
it has been proven that there exists a unique value $p_\rr{co}$ 
of the pressure, called \textit{coexistence pressure}, 
such that the potential energy of the two phases is equal.
More precisely, 
it has been proven that the equation 
$\Psi(m_\rr{s}(p),\varepsilon_\rr{s}(p))
 =\Psi(m_\rr{f}(p),\varepsilon_\rr{f}(p))$ has the single 
solution $p_\rr{co}$. 
We remark that it is not possible to provide explicit 
formulas for the phase field values and for the coexistence 
pressure. Indeed, these quantities do depend on the 
physical parameters of the model $\alpha$, $a$, and $b$. 
In \cite{CIS2010,CIS2011} we have proved (uniformly in these 
parameters) the scenario depicted above and we have provided 
a strategy for the numerical determination of all the interesting 
quantities. 

The behavior of the system at the coexistence pressure 
is particularly relevant; from now on we shall always 
consider $p=p_\rr{co}$ and, for this reason, we shall drop $p$ 
from the notation. 
When the external pressure is equal to $p_\rr{co}$,
none of the two above phases is 
favored and we ask if profiles connecting one phase to the other exist
\cite{CIS2011b}.
In \citep{CIS2010} this problem has been addressed on the 
space set $\bb{R}$, while some results on a finite interval 
have been reported in \cite{CIS2011b}. 
There we have remarked that 
the localization properties of the kink profile are not trivial, 
as it has been fully explained above.

The profiles connecting the two phases on a finite interval 
$[0,\ell]$ at the coexistence pressure 
are given by the solutions (if any) 
of the Dirichlet boundary problem for the Euler--Lagrange equations 
associated with the functional (\ref{funzionale02}), namely 
\begin{equation}
\label{el00}
\frac{\partial\Psi}{\partial\varepsilon}
-
\frac{\rr{d}}{\rr{d}x}\frac{\partial K}{\partial\varepsilon'}
=0
\;\textrm{ and }\;
\frac{\partial\Psi}{\partial m}
-
\frac{\rr{d}}{\rr{d}x}\frac{\partial K}{\partial m'}
=0
\end{equation}
(see, for instance, \cite[equation~(4)]{CIS2010}), with 
boundary conditions 
$m(0)=m_\rr{s}$,
$\varepsilon(0)=\varepsilon_\rr{s}$,
$m(\ell)=m_\rr{f}$,
and 
$\varepsilon(\ell)=\varepsilon_\rr{f}$.
By writing explicitly the Euler--Lagrange equations we finally 
get the Dirichlet problem
\begin{widetext}
\begin{equation}
\label{problema-staz00}
\left\{
\begin{array}{l}
k_1\varepsilon''+k_2m''
=
-(2/3)\alpha b m^3+\alpha b^2m^2\varepsilon+p+\varepsilon-ab(m-b\varepsilon)
\\
k_2\varepsilon''+k_3m'' 
=
 \alpha m^3-2\alpha b m^2 \varepsilon+\alpha b^2m\varepsilon^2
 +a(m-b\varepsilon)
\\
{\displaystyle
m(0)=m_\rr{s},\,
\varepsilon(0)=\varepsilon_\rr{s},\,
m(\ell)=m_\rr{f},\,
\varepsilon(\ell)=\varepsilon_\rr{f}
}
\\
\end{array}
\right.
\end{equation}
\end{widetext}
which can be approached numerically 
with the finite difference method powered with the Newton--Raphson algorithm.
The solution we find, in the cases of interest, are those depicted in 
the figures~\ref{f:orbe1}--\ref{f:orbm2}. 

As noted above the behavior at small second gradient 
coefficient $k_1$, $k_2$, and $k_3$, of the stationary profiles 
depicted in figures~\ref{f:orbe1} -- \ref{f:orbm2}
is very peculiar:
when $k_1,k_2,k_3\to0$ with 
mutual ratios kept constant, the interface becomes sharp and its location 
tends to a definite position not depending on these ratios. 
Although the general results in Section~\ref{s:risultati} do not apply 
to this case, 
we note that the solutions of the problem 
(\ref{problema-staz00}) behave in tune with those of the 
general one--field problem (\ref{dir00}).

At the moment, as already mentioned above,
we can explain this fact only in the particular case 
$k_1k_3-k_2^2=0$, namely, $k_1/k_2=k_2/k_3$.
In this case, called \textit{degenerate} case in \cite{CIS2010}, the 
problem of finding a solution of the stationary problem 
can be reduced to a one--field problem. 
Indeed, in such a case one performs the rotation
of the Cartesian reference system 
\begin{equation}
\label{secondo07}
\xi:=\frac{m+\lambda\varepsilon}{\sqrt{1+\lambda^2}}
\;\;\;\textrm{ and }\;\;\; 
\eta:=\frac{-\lambda m+\varepsilon}{\sqrt{1+\lambda^2}}
\end{equation}
in the plane $m$--$\varepsilon$, 
where $\lambda:=k_1/k_2=k_2/k_3$,
and defines 
\begin{equation}
\label{secondo08}
U(\xi,\eta)
=
\Psi(m(\xi,\eta),\varepsilon(\xi,\eta))
\end{equation}
Then one shows that the two fields 
$m(x)$ and $\varepsilon(x)$ are solutions of the two equations 
(\ref{problema-staz00}) if and only if the corresponding fields 
$\xi(x)$ and $\eta(x)$ satisfy
\begin{equation}
\label{secondo12}
k_3(1+\lambda^2)
\xi''=\frac{\partial U}{\partial \xi}(\xi,\eta)
\;\;\;\textrm{ and }\;\;\;
\frac{\partial U}{\partial \eta}(\xi,\eta)=0
\end{equation}
The root locus of 
the \textit{constraint curve}
$\partial U(\xi,\eta)/\partial\eta=0$ is made of 
a certain number of maximal components 
such that each of them is the graph of a 
function $\xi\in\bb{R}\to \eta(\xi)\in\bb{R}$;
for each of them the first between the two equations
(\ref{secondo12}) becomes
a one--field one--dimensional problem.

Since
the function $U$ has been obtained by 
rotating the coordinate axes,
then at the coexistence pressure it 
has the two absolute minimum points 
$(\xi_\rr{s},\eta_\rr{s})$
and 
$(\xi_\rr{f},\eta_\rr{f})$
corresponding, respectively, to the standard and to the fluid--rich phases. 
Since $(m_\rr{s},\varepsilon_\rr{s})$
and $(m_\rr{f},\varepsilon_\rr{f})$
satisfy the equations $\Psi_m(m,\varepsilon)=0$ and 
$\Psi_\varepsilon(m,\varepsilon)=0$, we have that the two points
$(\xi_\rr{s},\eta_\rr{s})$ and 
$(\xi_\rr{f},\eta_\rr{f})$ are solutions of the constraint equation 
$\partial U(\xi,\eta)/\partial\eta=0$
and hence they belong to the constraint curve. 

In \citep{CIS2010} we have seen that there exist
values of the second gradient parameters $k_1$, $k_2$, and $k_3$
such that the two points above 
fall on the same maximal component of the constraint equation.
Since, in this case, the function $U$ has two isolated absolute
minima which, by hypothesis, belong to the 
same maximal component of the 
constraint curve, we have that the function $U(\xi,\eta(\xi))$ of 
$\xi$ has two absolute isolated minima in 
$\xi_\rr{s}$ and $\xi_\rr{f}$. 

In this particular case, finally, the theory developed 
in Section~\ref{s:risultati} can be applied to 
the poromechanics problem (\ref{problema-staz00}) 
and the results discussed in the introduction find a complete
explanation. 
The explanation is not only qualitative, but also quantitative, 
indeed, for the first gradient parameters chosen as in the 
figures~\ref{f:orbe1} and \ref{f:orbm1},
namely, 
$\alpha=100$, $a=1/2$, and $b=1$, 
by computing the second derivatives of the one--field 
potential energy $U(\xi,\eta(\xi))$ 
at the phases $\xi_\rr{s}$ and $\xi_\rr{f}$ and 
by using (\ref{interfaccia01}) we get $0.6164$ for the position 
of the interface on the interval $[0,1]$. 
For the second gradient parameter choice in 
figures~\ref{f:orbe1} and \ref{f:orbm1}
the degeneracy condition $k_1k_3-k_2^2=0$ is satisfied. 
So that in this case the theory developed in Section~\ref{s:risultati}
applies and the match with the numerical result is striking. 

\section{Two numerical examples}
\label{s:motore}
In this section 
we check numerically (\ref{interfaccia00}) for two
very well known potentials.
The first example, see (\ref{adw01}), is an asymmetric  potential 
with mutually different second derivatives at the minima; the numerical 
computation will confirm that, for $\piccolo$ small,
the interface is not located at the middle 
point of the space interval
and that its position is given by (\ref{interfaccia01}).
The second example, see (\ref{adw02}), is, on the other hand, 
an asymmetric potential 
with mutually equal second derivatives at the minima; the numerical 
computation will confirm that, for $\piccolo$ small,
the interface is located at the middle 
point of the space interval.

We shall solve the equation (\ref{dir00}) with 
$\ell=1$ via the following procedure. 
Fix $\piccolo$ and use (\ref{disc02}) to compute 
the energy level $E_\piccolo$; then use (\ref{dir01}) 
to compute the profile. 

\begin{figure}
\begin{picture}(100,110)
\put(-70,-30)
{
\resizebox{14cm}{!}{\rotatebox{0}{\includegraphics{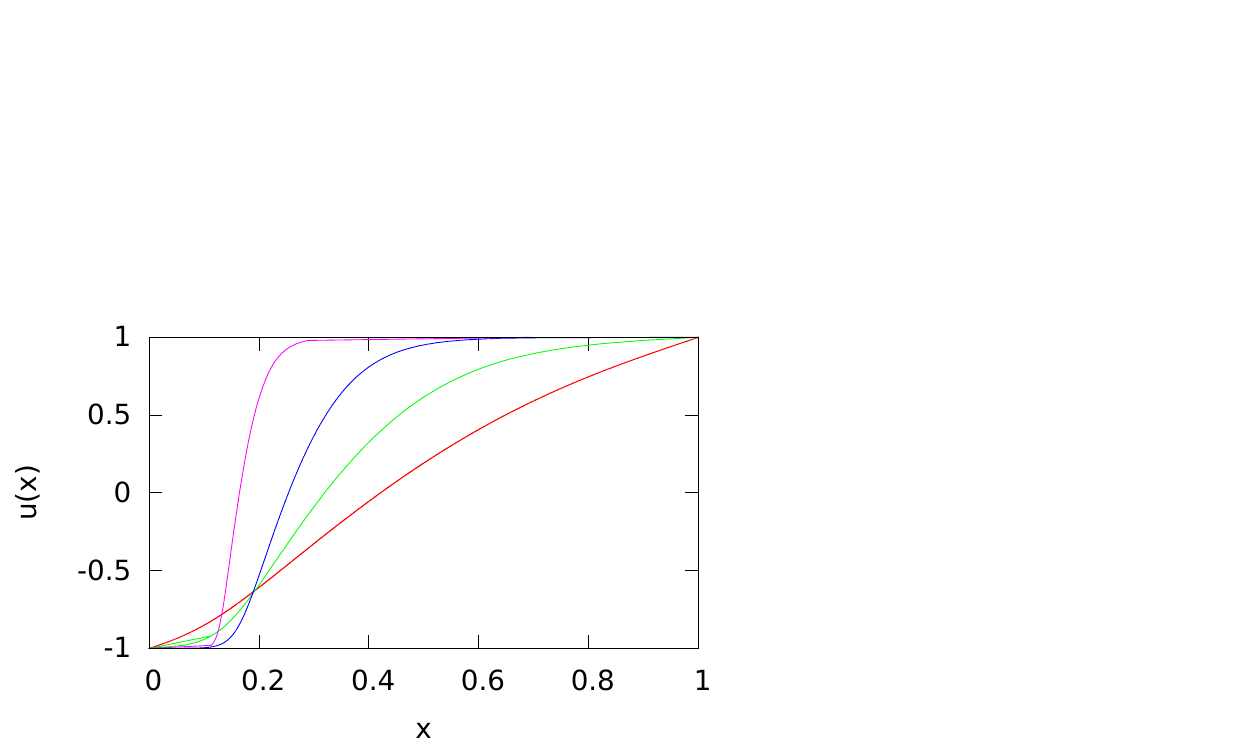}}}
}
\end{picture}  
\vskip 1. cm
\caption{Graph of the kink for the potential (\ref{adw01}) for 
$\omega_0=1$, $b_1=1$, $b_2=5$, and $u_0=0.5$.
The profiles correspond to 
$\piccolo=1.0,0.5,0.25,0.1$ 
in increasing order of steepness.
}
\label{f:orbv1}
\end{figure}

We consider first the following asymmetric double--well potential \cite{CM,STZ}
\begin{equation}
\label{adw01}
\begin{array}{l}
{\displaystyle
 V_1(u)=\frac{1}{\omega_0^2}
      \bigg(\frac{1-\exp[b_1(u-1)]}
                {1-\exp[b_1(u_0-1)]}
 \vphantom{\bigg\{_)\}}
}
\\
\phantom{
         V_1(u)=\frac{1}{\omega_0^2}
         \bigg(
        }
{\displaystyle
 \times
 \frac{1-\exp[-b_2(u+1)]}
      {1-\exp[-b_2(u_0+1)]}\bigg)^2
}
\\
\end{array}
\end{equation}
where $\omega_0\in\bb{R}$, $b_1,b_2>0$, and $u_0\in(-1,+1)$. 
The two minima are at $\pm1$ and 
$u_0$ controls the height of the internal barrier.
The kinks corresponding to $\piccolo=1.0,0.5,0.25,0.1$ 
(for $\omega_0=1$, $b_1=1$, $b_2=5$, $u_0=0.5$)
are shown in 
figure~\ref{f:orbv1}.
By applying (\ref{interfaccia01})
one gets 
$x^\rr{i}_0
=
[1+e^2+e^4+e^6+e^8]/[1+e^2+e^4+e^6+6e^8]
\approx
0.187846
$,
so that the numerical computation is in good agreement 
with the theoretical prediction (\ref{interfaccia01}).

\begin{figure}
\begin{picture}(100,110)
\put(-70,-30)
{
\resizebox{14cm}{!}{\rotatebox{0}{\includegraphics{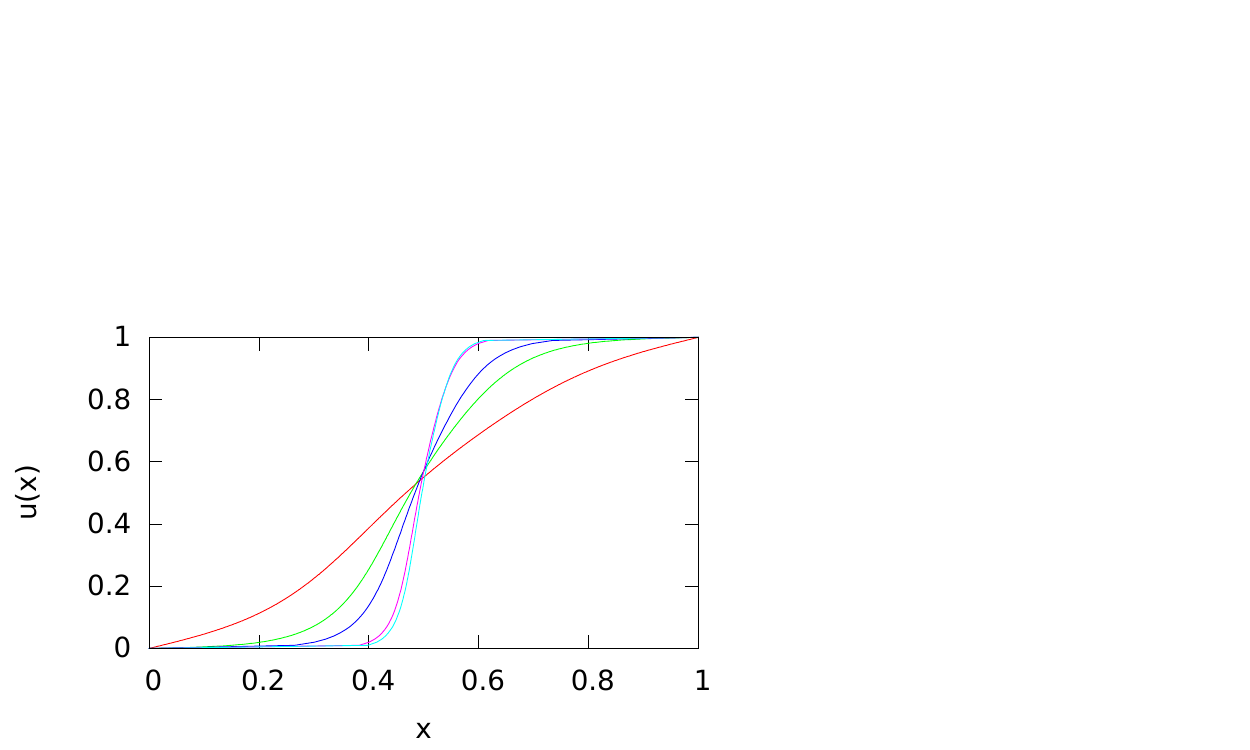}}}
}
\end{picture}  
\vskip 1. cm
\caption{Graph of the kink for the potential (\ref{adw02}) for 
$a=1$.
The profiles correspond to 
$\piccolo=3.0,1.5,1.0,0.5,0.45$
in increasing order of steepness.
The two profiles corresponding to $\piccolo=0.5$ and 
$\piccolo=0.45$ are almost coincident in the picture.
}
\label{f:orbv2}
\end{figure}

Another interesting example is the 
asymmetric potential which is 
considered the benchmark potential \cite{CMB,HM,Mateos}
of the current literature on rocked ratchets, namely, 
\begin{equation}
\label{adw02}
V_2(u)=
\sigma\sin\bigg[\frac{2\pi}{a}(u-u_0)\bigg]
+
\frac{\sigma}{4}\sin\bigg[\frac{4\pi}{a}(u-u_0)\bigg]
+R
\end{equation}
with $a\in\bb{R}$,
$\sigma^{-1}=(a/(8\pi))(\sqrt{3}/2)^{1/2}(3+\sqrt{3})$,
$u_0=(a/(2\pi))\cos^{-1}((-1+\sqrt{3})/2)$,
and 
$R=-2\pi/a$.
The not essential 
additive constant $R$ in (\ref{adw02}) has been introduced 
in order to ensure that the potential is equal to zero at the minima
$n a$ with $n\in\bb{Z}$. 
The second derivative of $V_2$ at the minima $na$ are all
equal since the function is periodic with period $a$. 
Some algebra yields
$V''_2(na) = 16(-1+\sqrt{3})\pi^3/a^3$, so that these derivatives 
are not zero.
The kinks corresponding to 
$\piccolo=3.0,1.5,1.0,0.5,0.45$ (for $a=1$)
are shown in 
figure~\ref{f:orbv2}.

We stress that, in this case, although the potential energy is asymmetric 
with respect to the phase exchange, the interface tends to be located 
in the middle point of the space interval, since the second derivatives 
of the potential energy at the minima are mutually equal. 

Note that our result applies even if deformations of the 
standard rocket ratchet potential (\ref{adw02}) are considered 
(see \cite{AVM}). For instance 
the positive multiplicative factor $1/4$ in the second term in (\ref{adw02}) 
can be chosen differently. Our result is still valid, provided 
no other minima is found between $0$ and $a=1$, since the 
second derivatives of the potential at the minima are equal and do not vanish. 

It is notable to remark the following. 
In the above sections we have studied the localization properties 
of the interfaces connecting two phases under an 
asymmetric double--well potential energy with degenerate minima and we have 
stated (see Section~\ref{s:dimostrazione} for the proof)
that the position of the interface depends 
exclusively on the second derivatives of the potential 
energies at the two phases. 

Although
we cannot say, at the moment, to which extent this 
analogy between the two results can be pushed forward,
we note that our result is similar in spirit to those related 
to the direction of motion of a dc soliton current in 
ac drive rocket ratchet.
Which is, a priori, a completely different problem 
with respect to the phase interface problem (\ref{dir00}) that 
we have studied. 

We refer in particular to \cite{CM,STZ}. 
There the authors considered the potential (\ref{adw01}) 
and proved the following: a string soliton profile is pushed by 
``symmetric" thermal fluctuations sidewise towards the 
boundary point where the boundary condition has been chosen equal to 
the phase corresponding to the narrowest valley (biggest
second derivative) of the double--well
potential.
In particular in \cite{STZ} it is shown that this tendency 
to move towards the narrowest valley is controlled by the 
second derivatives of the potential energy evaluated at the phases. 

\section{Proof of results}
\label{s:dimostrazione}
The proof of the results in Section~\ref{s:risultati} is based on 
the classical Weierstrass qualitative analysis of the 
one--dimensional mechanical problem equivalent to (\ref{dir00}) \cite{CI}.

Before starting the proof of the results, 
we note that $E_\piccolo$ is an increasing positive function 
of $\piccolo$ and $E_\piccolo\to0$ for $\piccolo\to0$. 
This is a quite intuitive result since taking the limit $k\to0$ 
is the same as considering a zero mass limit in the equivalent mechanical 
system. 
More precisely, 
remark that 
\begin{displaymath}
\int_a^b\frac{\rr{d}s}
             {\sqrt{2[E+V(s)]}}
\end{displaymath}
is an decreasing function of $E$, defined for $E>0$, and such that
its limit for $E\to0$ tends to infinity.
Since
by (\ref{disc02}) 
\begin{displaymath}
\int_a^b\frac{\rr{d}s}
             {\sqrt{2[E_\piccolo+V(s)]}}
=
\frac{\ell}{\piccolo}
\end{displaymath}
it follows that for $\piccolo\to0$ the energy 
level $E_\piccolo$ tends to $0$.

The Theorem~\ref{t:disc} is a classical result 
in interface theory. 
For the sake of completeness we report the proof in our case.

\smallskip 
\textit{Proof of Theorem~\ref{t:disc}.\/}
Since $V$ is a strictly positive function in 
$[u_1,u_2]$, by the dominated convergence theorem it follows immediately that 
\begin{displaymath}
\lim_{\piccolo\to0}
\int_{u_1}^{u_2}
\frac{\rr{d}s}
             {\sqrt{2[E_\piccolo+V(s)]}}
=
\int_{u_1}^{u_2}
\frac{\rr{d}s}
             {\sqrt{2V(s)}}
<\infty
\end{displaymath}
yielding (\ref{disc}).
\qed

The proof of Theorem~\ref{t:interfaccia00}, based on 
the heuristic arguments discussed in Section~\ref{s:risultati}, uses
some ideas developed in \cite[Appendix~A]{BBB}.

\smallskip 
\textit{Proof of Theorem~\ref{t:interfaccia00}.\/}
Recall the definition (\ref{interk}) of $\interk$.
By simple algebra we have that 
\begin{equation}
\label{dimo00}
\interk
=
\piccolo\,S_{a,\piccolo}+\piccolo\,R_{a,\piccolo}
\end{equation}
where 
\begin{displaymath}
S_{a,\piccolo}
=
\int_a^{\frac{a+b}{2}}
\frac{\rr{d}s}
             {\sqrt{2[E_\piccolo+V''(a)(s-a)^2/2]}}
\end{displaymath}
and  
\begin{displaymath}
R_{a,\piccolo}
=
\int_a^{\frac{a+b}{2}}
\frac{\rr{d}s}
             {\sqrt{2[E_\piccolo+V(s)]}}
-S_{a,\piccolo}
\end{displaymath}
By direct integration we get 
\begin{equation}
\label{dimo02}
S_{a,\piccolo}
=
\frac{1}{\sqrt{V''(a)}}
  \,
  \mathrm{arcsinh}\bigg[
               \sqrt{\frac{V''(a)}{2E_\piccolo}}
               \,\frac{b-a}{2}
          \bigg]
\end{equation}

The idea behind the proof is the following: once 
$S_{a,\piccolo}$ has been 
computed explicitly,
we are able to manage the divergence 
of the integral with $E_\piccolo$ for $\piccolo\to0$. 
On the other hand we shall simply estimate $R_{a,\piccolo}$ 
and prove that it is finite for any $k\ge0$. 

We can perform the analogous computation in the neighborhood 
of the boundary point $\ell$ where the boundary condition $b$ 
has been fixed. 
Noted that
\begin{displaymath}
\ell-\interk
=
\int_{\frac{a+b}{2}}^b
\frac{\piccolo\,\rr{d}s}
             {\sqrt{2[E_\piccolo+V(s)]}}
\end{displaymath}
we get 
\begin{equation}
\label{dimo01}
\ell-\interk
=
\piccolo\,S_{b,\piccolo}+\piccolo\,R_{b,\piccolo}
\end{equation}
where 
\begin{equation}
\label{dimo03}
\begin{array}{rcl}
S_{b,\piccolo}
&\!\!=&\!\!
{\displaystyle
\int_{\frac{a+b}{2}}^b
\frac{\rr{d}s}
             {\sqrt{2[E_\piccolo+V''(b)(b-s)^2/2]}}
\vphantom{\bigg\{_\bigg\}}
}
\\
&\!\!=&\!\!
{\displaystyle
\frac{1}{\sqrt{V''(b)}}
  \,
  \mathrm{arcsinh}\bigg[
               \sqrt{\frac{V''(b)}{2E_\piccolo}}
               \,\frac{b-a}{2}
          \bigg]
}
\end{array}
\end{equation}
and  
\begin{displaymath}
R_{b,\piccolo}
=
\int_{\frac{a+b}{2}}^b
\frac{\rr{d}s}
             {\sqrt{2[E_\piccolo+V(s)]}}
-S_{b,\piccolo}
\end{displaymath}

Recalling that $\mathrm{arcsinh}(x)=\log(x+\sqrt{x^2+1})$, 
by equations (\ref{dimo00})--(\ref{dimo03}) we get 
\begin{widetext}
\begin{equation}
\label{dimo04}
\sqrt{V''(a)}\interk-\sqrt{V''(b)}(\ell-\interk)
=
\piccolo
\bigg[
\log\frac
       {\sqrt{V''(a)}(b-a)+\sqrt{V''(a)(b-a)^2+8E_\piccolo}}
       {\sqrt{V''(b)}(b-a)+\sqrt{V''(b)(b-a)^2+8E_\piccolo}}
+\sqrt{V''(a)}R_{a,\piccolo}
-\sqrt{V''(b)}R_{b,\piccolo}
\bigg]
\end{equation}
\end{widetext}
which is the analogous of the equation (\ref{heur00}) that 
we obtained in the heuristic discussion in Section~\ref{s:risultati}.

By using the explicit expression of $S_{a,\piccolo}$ and 
$S_{b,\piccolo}$ we have been able to cancel the divergence in 
$E_\piccolo$ in 
the first part of the right--hand side of (\ref{dimo04}).
Recall equation (\ref{interfaccia00});
since $E_\piccolo\to0$ for $\piccolo\to0$, 
the theorem will follow
from (\ref{dimo04}) once we shall have proven that 
\begin{equation}
\label{dimo05}
\sup_{\piccolo\ge0}|R_{a,\piccolo}|<\infty
\;\;\;\textrm{ and }\;\;\;
\sup_{\piccolo\ge0}|R_{b,\piccolo}|<\infty
\end{equation}

We prove the first of the two bounds above; the second one can be 
proven similarly. With some algebra we get
\begin{widetext}
\begin{displaymath}
\begin{array}{rcl}
R_{a,\piccolo}
&\!\!=&\!\!
{\displaystyle
 \int_a^{\frac{a+b}{2}}
 \frac
  {\sqrt{2E_\piccolo+V''(a)(s-a)^2}-\sqrt{2[E_\piccolo-V(s)]}}
  {\sqrt{2E_\piccolo+V''(a)(s-a)^2}\sqrt{2[E_\piccolo-V(s)]}}
 \,\rr{d}s
 \vphantom{\bigg\{_\bigg\}}
}
\\
&\!\!=&\!\!
{\displaystyle
\int_a^{\frac{a+b}{2}}
 \frac{[-2V(s)+V''(a)(s-a)^2]\,\rr{d}s}
      {\sqrt{2[E_\piccolo+V(s)]}[2E_\piccolo+V''(a)(s-a)^2]
       +
       2[E_\piccolo+V(s)]\sqrt{2E_\piccolo+V''(a)(s-a)^2}
      }
}
\end{array}
\end{displaymath}
\end{widetext}
where in the last step we have multiplied numerator and denominator 
times the sum of the two square root terms.
Since, as we remarked at the beginning of this section,
$E_\piccolo>0$ for any $\piccolo>0$ and 
$E_\piccolo\to0$ for $\piccolo\to0$, we have that 
\begin{widetext}
\begin{displaymath}
\sup_{\piccolo\ge0}|R_{a,\piccolo}|
\le
\int_a^{\frac{a+b}{2}}
 \frac{|-2V(s)+V''(a)(s-a)^2|\,\rr{d}s}
      {2V(s)\sqrt{V''(a)}(s-a)
       +
       \sqrt{2V(s)}V''(a)(s-a)^2
      }
\end{displaymath}
\end{widetext}
Moreover
the integrand is continuous in $(a,(a+b)/2]$ and 
its limit for $s\to a$ is $|V'''(a)|/[6V''(a)\sqrt{V''(a)}]$; 
therefore, we have that
the integral on the right--hand side of the inequality above is finite. 
We thus get the first of the two bounds (\ref{dimo05}).

As noted below (\ref{dimo04}) this is sufficient to 
complete the proof of the theorem.
\qed

\begin{acknowledgements}
We are indebted to 
E.\ Presutti for having suggested a 
preliminary conjecture on which these results are based
and to 
F.\ Marchesoni for many valuable comments and suggestions. 
The authors also thank P.\ Butt\`a and L.\ Bertini for useful discussions.
\end{acknowledgements}


\end{document}